\documentclass[12pt]{JHEP3}
\usepackage{epsfig}

\def\zzz{{\hbox{z\kern-1mm z}}}

\newcommand{\be}{\begin{equation}}
\newcommand{\ee}{\end{equation}}
\newcommand{\ben}{\begin{eqnarray}}
\newcommand{\een}{\end{eqnarray}}
\newcommand{\ba}{\begin{eqnarray}}
\newcommand{\ea}{\end{eqnarray}}
\newcommand{\nn}{\nonumber \\}

\newcommand{\beq}{\begin{equation}}
\newcommand{\eeq}{\end{equation}}

\newcommand{\bi}{\begin{itemize}}
\newcommand{\ei}{\end{itemize}}
\newcommand{\ii}{\item}

\newcommand{\lb}{\left(}
\newcommand{\rb}{\right)}

\newcommand{\bc}{\begin{center}}
\newcommand{\ec}{\end{center}}

\title{Subleading Correction to Statistical Entropy for BMPV Black
    Hole\footnote{Based on talks given at LPTHE, Jussieu, AEI, Potsdam, ETH,
    Zurich, INFN Rome.}}

\author {Nabamita Banerjee \\
{\it Harish-Chandra Research Institute, Chhatnag Road, Jhunsi,
  Allahabad 211 019, INDIA}\\
Email: {\tt nabamita@mri.ernet.in}}
\abstract{We study higher derivative corrections to the statistical
  entropy function and the statistical entropy for five dimensional
  BMPV black holes by doing the asymptotic expansion of the partition
  function. This enables us to evaluate entropy for a large range of charges,
  even out of Cardy (Farey tail) limit. }


\preprint{}

\begin{document}

\section{Introduction } \label{intro}

Counting of 1/4 BPS dyonic states in four dimensional ${\cal N}=4$
supersymmetric string theories has been studied in great detail in last few
years\cite{dvv},\cite{sen1},\cite{sen2},\cite{banerjee:2007ub},\cite{atish1},\cite{shamik1},\cite{shamik2},\cite{sameer1}.
We now have a good understanding of the degeneracy formula, its moduli
dependence and the wall crossing formulae.  Large charge asymptotic expansion
of these degeneracy formulae exactly capture the dyonic black hole entropy
including certain subleading corrections due to higher derivative corrections
to the supergravity.

Five dimensional spinning (BMPV) black holes\cite{bmpv} is a close cousin of
the four dimensional dyonic black hole.  These black holes were first
constructed in \cite{bmpv}, as a spinning generalization of \cite{sv}. These
are charged, spinning, 5-dimensional black holes with constant dilation and
constant moduli in type $IIB$ theory on $K3 \times S^1$. The microscopic
configurations of these black holes can be described as p-solitonic states
(Dp- branes) in type IIB theory on $K3 \times S^1$, for $p=1,3,5$. The states
also carry certain momenta along the $S^1$ circle and angular momenta along
the non compact directions. This microscopic description is much similar to
that of the four dimensional dyonic black holes. In fact,when described in
terms of D-branes, the BMPV black hole consists of D1-D5-p system, whereas the
four dimensional dyonic black hole in addition has a KK monopole background.
It is therefore natural to study BMPV black hole entropy in terms of the four
dimensional dyon degeneracy formula without KK monopole contribution.  A
general degeneracy formula for D1-D5-p system is then easy to write
down. However, we are interested in finding out subleading correction to the
BMPV black hole entropy due to higher derivative terms in the effective
action. Higher derivative correction to five dimensional black holes has been
computed in \cite{sm} and their entropy has been computed \cite{castro}. In
this paper we will take a different approach to this problem.  Determination
of subleading correction is done in a most effective fashion using the
statistical entropy function and the effective action formalism.  Using the
statistical entropy function one can write down a one dimensional effective
action, and using the Feynman diagram technique one can obtain systematic
large charge asymptotic expansion of the statistical entropy.  This method
correctly reproduces subleading correction to the entropy of four dimensional
1/4 BPS dyonic black holes.

This particular feature of the statistical entropy function gives the
motivation to compute similar subleading correction to the five
dimensional BMPV black hole. The exactness of the statistical entropy
(or the statistical entropy function) suggests that we can evaluate
the entropy (or the entropy function) to any order.

The rests of the paper is divided into three sections. In the first section,
we present a different form of the degeneracy function of the five dimensional
BMPV black holes, based on the degeneracy of the four dimensional dyonic black
holes. In the next section we discuss the first subleading $(O(Q^0))$
correction to the statistical entropy function and statistical entropy of
these black holes. In the last section, we have some discussions on our
results.

As this paper was being written a paper \cite{sameer3} appeared in the arXiv
that discusses similar issues.

\section{Degeneracy Function For 5-dimensional BMPV Black Holes }

In this section, we rewrite the degeneracy function of the BMPV black
holes in a different form. The microscopic description for this black
hole is a particular D-brane configuration in type IIB theory
compactified on $K3 \times S^1$.  This contains $Q_1$ number of $D1$
branes, $Q_5$ number of D5 branes, $-n$ units of
momenta along $S^1$ circle and angular momenta J1 and J2 along the
non-compact spatial directions.  This configuration, however does not
contain any D3 branes. For extremal black holes, the
corresponding microscopic configuration requires the modulus of the
two angular momenta to be same $|J1|=|J2|=J$. The microscopic
computation for the leading entropy was first done in \cite{bmpv}. We
will write $\alpha'$ (inverse string tension) exact degeneracy
function for this configuration, from the knowledge of the degeneracy
function of 4-dimensional dyonic black holes in ${\cal N}=4$
supersymmetric string theory. Here we will sketch in brief how the
degeneracy function was obtained for these 4-dimensional black holes
\cite{justin},\cite{review}.

\subsection{Degeneracy Function of 4-dimensional Dyonic Black holes}

Let us consider type IIB theory compactified on $K3 \times S^1 \times \tilde
S^1$. Following a chain of duality transformations, one can look at the same
theory as a heterotic string theory compactified on $T^6$. These theories have
dyonic black-hole solutions.  Let us consider a specific configuration in this
compactified type IIB theory : $Q_1$ number of D1-branes wrapped along $S^1$,
$Q_5$ number of D5-brane wrapped along $K3 \times S^1$, a single Kaluza-Klein
monopole associated with $\tilde S^1$ circle,$ -n$ units of momentum along
$S^1$ direction and J units of angular momentum along $\tilde S^1$
direction. In the dual Heterotic picture, this represents dyonic black hole
solutions. If we stay in a region of the moduli space where the type IIB
theory is weakly coupled, the partition function of the entire system can be
obtained by considering three weakly interacting sources:

\begin{enumerate}
\item the relative motion of the D1-brane in the plane of
D5-brane, carrying certain momenta $-L$ along $S^1$ and $J'$ $\tilde
S^1$ directions,
\item the center of mass motion of D1-D5 system in
the KK-monopole background carrying momenta $-l_0$ along $S^1$
and $j_0$ along $\tilde S^1$ directions,
\item excitations of the
KK-monopole carrying $-l_0'$ momentum along $S^1$, 
\end{enumerate}
with $n=L+l_0+l'_0$ and $J=J'+j_0$ being the sum of momenta along
$S^1$ and $\tilde S^1$ directions respectively. Hence, in the weak
coupling limit, the partition function $f(\tilde
\rho,\tilde\sigma,\tilde v )$ of the configuration can be expressed
as,

\begin{eqnarray}
f(\tilde \rho,\tilde \sigma,\tilde v ) &=& 
-{1\over 64} \, \left(
\sum_{Q_1,L,J'} (-1)^{J'} \, d_{D1}(Q_1,L,J') 
e^{2\pi i ( 
\tilde \sigma Q_1 /N +\tilde \rho L + \tilde v J')}\right) \nn
&& \, \left(\sum_{l_0, j_0} (-1)^{j_0}
d_{CM}(l_0, j_0) e^{2\pi i l_0\tilde\rho + 2\pi i
j_0\tilde v}\right) \, 
\left(\sum_{l_0'} d_{KK}(l_0') e^{2\pi i l_0' \tilde \rho} \right)\, ,
\end{eqnarray}
where $d_{D1}(Q_1,L,J')$ is the degeneracy of source $(1)$,
$d_{CM}(l_0,j_0)$ is the degeneracy associated with source $(2)$ and
$d_{KK}(l_0')$ denotes the degeneracy associated with source
$(3)$. The factor of 1/64 accounts for the fact that a single quarter
BPS supermultiplet has 64 states. Evaluating these three pieces
separately, the full partition function of the system looks like,

\begin{equation}\label{deff}
f(\tilde \rho,\tilde \sigma,\tilde v )= e^{-2\pi i (\tilde \rho 
+ \tilde v)}\prod_{k'\in \zzz+{r},l\in \zzz, j\in 2\zzz
\atop k',l\ge 0 , j<0 \, {\rm for}
\, k'=l=0}
\left( 1 - e^{2\pi i (\tilde \sigma k'   + \tilde \rho l + \tilde v j)}
\right)^{- c(4lk' - j^2)}\,. 
\end{equation}

Then we define the degeneracy function $ \tilde\Phi(\tilde
\rho,\tilde \sigma, \tilde v)$ and degeneracy of states $d(\vec Q,
\vec P)$ as, 
\begin{eqnarray}\label{phidqp} 
f(\tilde \rho,\tilde \sigma,\tilde v )
&=& {e^{2\pi i \tilde \sigma } \over \tilde\Phi(\tilde \rho,\tilde
  \sigma, \tilde v)},\, \nn 
d(\vec Q, \vec P) &=& (-1)^{Q\cdot P+1}\,
h\left({1\over 2} Q^2 , {1\over 2}\, P^2, Q\cdot P\right)\, , 
\end{eqnarray}
where $(\vec Q, \vec P)$ are the charge vectors carried by the black holes:

\be
Q= \
\lb 
\begin{array}{c}
0\\
-n\\
0\\
 -1
\end{array} \rb,\ \ \ \ \ \ \ \ 
P= \
\lb
\begin{array}{c}
Q_5(Q_1-Q_5)\\
-J\\
Q_5\\
0
\end{array}\rb.
\ee
and $h(m,n,p)$ are the coefficients of Fourier expansion of the
function $1/ \tilde\Phi(\tilde \rho,\tilde \sigma, \tilde v)$:
\begin{equation}\label{efo2} 
{1 \over \tilde \Phi(\tilde \rho,\tilde \sigma, \tilde
  v)} =\sum_{m,n,p} g(m,n,p) \, e^{2\pi i (m\, \tilde \rho + n\,
  \tilde\sigma + p\, \tilde v)}\, .  
\end{equation}

At this point it is worth noting from equations (\ref{deff}) and
(\ref{efo2}) that the power series gets a contribution $e^{-2 \pi i
\tilde v}(1-e^{-2 \pi i \tilde v})^{-1}$ from $k'=l=0$ term and one
can expand the series either in $e^{-2 \pi i \tilde v}$ or in $e^{2
\pi i \tilde v}$. There is an ambiguity in the expansion, we will come
back to this point in the last section .\\ 

To evaluate the degeneracy of a state associated with charges $(\vec
Q, \vec P)$, we need to invert equation (\ref{efo2}) as,
\begin{equation}\label{ehexp}
d(\vec Q, \vec P) =  (-1)^{ Q\cdot P+1}\, 
 \int _{\cal C} d\tilde \rho \, d\tilde\sigma \,
d\tilde v \, e^{- \pi i ( \tilde \rho Q^2 
+ \tilde \sigma P^2 +2 \tilde v Q\cdot P)}\, {1
\over \tilde \Phi(\tilde \rho,\tilde \sigma, \tilde v)}\, ,
\end{equation}
where ${\cal C}$ is a three real dimensional subspace of the three
complex dimensional space labeled by $(\tilde \rho,\tilde \sigma,
\tilde v)$, given by
\begin{eqnarray}\label{ep2int}
\tilde \rho_2=M_1, \quad \tilde\sigma_2 = M_2, \quad
\tilde v_2 = -M_3, \nonumber \\
 0\le \tilde\rho_1\le 1, \quad
0\le \tilde\sigma_1\le 1, \quad 0\le \tilde v_1\le 1\, .
\end{eqnarray}
$M_1$, $M_2$ and $M_3$ are large but fixed 
positive numbers with $M_3<< M_1, M_2$.
The choice of the $M_i$'s is determined from the requirement that
the Fourier expansion is convergent in the region of integration.

The ${\cal N}=4$ supersymmetric string theories discussed above are
invariant under $O(6,22, \mathbb{Z})$ T-duality and $ SL(2,\mathbb{Z})$ 
S-duality symmetry. The T-duality invariants are given as,
\be
Q^2=2n, \ \ \ \ P^2=2Q_5(Q_1-Q_5), \ \ \ \ \ \ Q\cdot P=J.
\ee
The function $\tilde \Phi$ actually behaves as a
modular form of weight $k=10$ under the S-duality group
$SL(2,\mathbb{Z})$.

\subsection{Degeneracy for BMPV Black Holes}

Here we will compute the degeneracy function for BMPV black holes from our
knowledge of the degeneracy function of 4-dimensional black holes we studied
in the last section. Comparing with the four-dimensional black hole, we will
treat the microscopic configuration of the BMPV black holes to be same as the
one considered in 4-dimensional case except for the following changes. The
radius of the $\tilde S^1$ circle is infinite and therefore the KK-monopole
sector is replaced by $\mathbf{R^4}$. We will again work in a region of the
moduli space where IIB theory is weakly coupled. The partition function of
this configuration will only get contribution from the source $(1)$ of
section$ (2.1)$, i.e., the relative motion of the D1-branes in the plane of
D5-branes. Hence, we have,
\begin{eqnarray} 
f_{bmpv}(\tilde \rho,\tilde \sigma,\tilde v )&=& -{1\over 64} \, \left(
\sum_{Q_1,L,J'} (-1)^{J'} \, d_{D1}(Q_1,L,J') e^{2\pi i ( \tilde
  \sigma Q_1 /N +\tilde \rho L + \tilde v J')}\right) \nn
&=&\prod_{k'\in \zzz,l\in \zzz, j\in 2\zzz \atop k'> 0, l \geq 0 }
\left( 1 - e^{2\pi i (\tilde \sigma k' + \tilde \rho l + \tilde v j)}
\right)^{- c(4lk' - j^2)} \ .
\end{eqnarray} 
Following the steps given in equations (\ref{phidqp}), (\ref{efo2})
and (\ref{ehexp}), we define the degeneracy function and degeneracy of
states for the BMPV black hole. The degeneracy function is given as,
\begin{eqnarray}
\tilde \Phi_{bmpv}(\tilde \rho, \tilde \sigma,\tilde v)&=&
{\tilde \Phi(\tilde \rho, \tilde \sigma,\tilde v) \over 
G(\tilde \rho, \tilde v)}, 
\een
where,
\ben \label{defg} 
  G(\tilde \rho, \tilde v)=64 e^{2 \pi i (\tilde \rho+
\tilde v)} (1-e^{-2 \pi i \tilde v})^2 \prod_{n=1}^{\infty}(1&-&e^{2 \pi i 
n \tilde \rho})^{20}(1-e^{2 \pi i(n \tilde \rho + \tilde v)})^2 \nn
(1&-&e^{2 \pi i(n \tilde \rho - \tilde v)})^2.
\end{eqnarray}
Here the function $G(\tilde \rho, \tilde v)$ basically captures the degeneracy
of the KK-monopole sector and the center of mass motion of the D1-D5 system in
KK-monopole background for four dimensional dyonic black holes.

\section{Correction to The Statistical Entropy Function}
Similar to the 4-dimensional black hole , we define the degeneracy of states
for the BMPV black holes as,
\be
\label{entropy}
d(\vec Q, \vec P)=  (-1)^{ Q\cdot P+1}\, 
 \int _{\cal C} d\tilde \rho \, d\tilde\sigma \,
d\tilde v \, e^{- \pi i ( \tilde \rho Q^2 
+ \tilde \sigma P^2 +2 \tilde v Q\cdot P)}\, {1
\over \tilde \Phi_{bmpv}(\tilde \rho,\tilde \sigma, \tilde v)} \ .
\ee 

The statistical entropy for the system is then given as, \be
S_{stat}=\ln d(\vec Q,\vec P) \ .  \ee One can evaluate the integral
(\ref{entropy}) by saddle point method and estimate the statistical
entropy for the system. We will take a different approach to estimate
the entropy. From the integral (\ref{entropy}), we will first evaluate
a function $\Gamma^{stat}$ analogous to black hole entropy
function. This function is called the statistical entropy function. The
statistical entropy is then obtained as the value of this function at
its extrema.  This can be done by following two steps:

\bi
\ii
The $v$ integral is done by residue methods. The function $\Phi_{bmpv}(\tilde 
\rho,\tilde \sigma, \tilde v)$ has a zero at 
\be \label{ezeropos}
\tilde \rho \tilde \sigma-\tilde v^2+ \tilde v=0.
\ee
Near this pole the function $\Phi_{bmpv}$ behaves as,
\be \label{ephibehav}
\Phi_{bmpv}(\tilde \rho,\tilde \sigma,\tilde v)
= (2v-\rho-\sigma)^{k} \, v^2 \, {g(\rho) \, g(\sigma) \over \hat G(\rho,
 \sigma, v) }\, ,
\ee
where 
\begin{equation}\label{e5nrep}
\rho 
   = {\tilde \rho \tilde \sigma - \tilde v^2\over \tilde \rho}, 
   \qquad \sigma = {\tilde \rho \tilde \sigma - (\tilde v - 1)^2\over  
   \tilde \rho}, \qquad
   v 
=   {\tilde \rho \tilde \sigma - \tilde v^2 + \tilde v\over \tilde \rho}\, ,
\end{equation}
$k$ is related to the rank $r$ of the gauge group via the
relation
\be \label{erank}
r = 2k + 8\, ,
\ee
and $g(\tau)$ is a known function which depends on the details
of the theory. Typically it transforms as a modular function of weight
$(k+2)$ under a certain subgroup of the $SL(2, \mathbb Z)$ group.
In the $(\rho,\sigma,v)$ variables the pole at (\ref{ezeropos})
is at $v=0$.  Near this pole, the integrand looks like, $v^{-2}F(\rho,\sigma,v)
 \hat G(\rho,\sigma,v)$ where,
\ben \label{def}
F(\rho,\sigma,v)&=&{(2 v - \rho - v)^{(-k-3)}\over g(\rho) g(\sigma)} 
   e^{\left [-i \pi \lb {v^2 -\rho
    \sigma \over 2v - \rho - \sigma} P^2 + {Q^2 \over 2v - \rho \sigma} + {2(v
    - \rho) \over 2v - \rho \sigma} Q\cdot P \rb \right]}. \nn
\hat G(\rho,\sigma,v)&=& 64 e^{2 \pi i {1+v-\rho \over 2v-\rho-\sigma}}
(1-e^{-2 \pi i {v-\rho \over  2v-\rho-\sigma}})^2\nn
&& \prod_{n=1}^{\infty}(1-e^{2 \pi i {n \over  2v-\rho-\sigma}})^{20}(1- e^{2 \pi i
  {
n+(v-\rho) \over  2v-\rho-\sigma}})^2 (1- e^{2 \pi i
  {
n-(v-\rho) \over  2v-\rho-\sigma}})^2 \ .
\een

After doing the $v$ integral using the above relation, (\ref{entropy}) 
takes the form, 
\be
\label{ek1}
e^{S_{stat}(\vec Q, \vec P)} \equiv d(\vec Q, \vec P)\simeq
\int{d^2\tau\over \tau_2^2} \, e^{-F_{bmpv}(\vec \tau)}\, ,
\end{equation}
where $\tau_1$ and $\tau_2$ are two complex variables, related
to $\rho$ and $\sigma$ via
\be \label{ereln}
\rho\equiv  \tau_1+ i \tau_2 \ \ ,\ \ \sigma\equiv \tau_1- i \tau_2\, ,
\ee
and the effective action $F_{bmpv}$ is given as,
\begin{eqnarray}\label{ek2}
F_{bmpv}(\vec \tau)&=&F(\vec \tau)-\ln \hat G(\vec \tau)-\ln \lb1+{f \hat G'
 \over \hat G f'}(\vec \tau)\rb \nn
F(\vec\tau) &=& -\Bigg[ {\pi\over 2 \tau_2} \, |Q -\tau P|^2
-\ln g(\tau) -\ln g(-\bar\tau) - (k+2) \ln (2\tau_2)
\nonumber \\
&& +\ln\bigg\{K_0 \, \left(
2(k+3) + {\pi\over \tau_2} |Q -\tau P|^2\right)
\bigg\}\Bigg] \, , \nonumber \\
K_0 &=& constant\, .
\end{eqnarray}
The function $F(\vec \tau)$ is actually the effective action for 4-dimensional
black holes. The function $\hat G(\vec \tau)$ and $f(\vec \tau)$ are same as $
\hat G( \rho, \sigma,v)$ and $ F( \rho, \sigma,v)$ in (\ref{def})
respectively, evaluated at $v=0$ and expressed as functions of $\vec
\tau$. Here $'$ means derivative with respect to $ v$ evaluated at $v=0$. We
give the expressions for the function $\hat G(\vec \tau)$ here for later use:
\begin{eqnarray} 
\hat G(\vec \tau)&=& -64 e^{-{\pi \over \tau_2}(1-\tau_1)}(1+e^{{-\pi
    \over \tau_2} \tau_1})^2\prod_{n=1}^{\infty}(1-e^{- n \pi \over
  \tau_2})^{20}(1+e^{{- \pi \over \tau_2}(n+\tau_1)})^2 (1+e^{{- \pi
    \over \tau_2}(n+\tau_1)})^2. \nn 
\end{eqnarray}
\ii 

Next we evaluate (\ref{ek1}) by considering it to be a zero
dimensional field theory with
fields $\tau, \bar \tau$ (or equivalently $\tau_1, \tau_2$) and action
$F_{bmpv}(\vec \tau)-2 \ln \tau_2$. We apply background field method
technique to obtain the statistical entropy function. In this method, we do
an asymptotic expansion of the action around a fixed background point $\vec
\tau_B$, which is not the saddle point of the action. This expansion is valid
for 
\be \label{limit} 
Q^2 >0 \ \ \ \ \ \ P^2>0\ \ \ \ \ \ \ \ \ Q^2 P^2>(Q\cdot P)^2\ .
\ee 

The statistical entropy function (to a certain order in charges) is
then given as a sum of all 1PI vacuum diagrams (required to that order)
in this zero dimensional field theory.
\ei

We now want to evaluate the four derivative, i.e., $O(Q^0)$ correction to the
statistical entropy. The last term in (\ref{ek2}) is of $O(Q^{-2n},
n\ge 1)$. Similarly, the last term in $F(\vec \tau)$ also goes as $O(Q^{-2n},
n\ge 1)$. Hence, up to order $Q^0$, these terms will not contribute. The first
term in $F(\vec \tau)$ is $O(Q^2)$, while the second term of $F(\vec \tau)$
and $F_{bmpv}(\vec \tau)$ are $O(Q^0)$. Therefore the first term needs to be
expanded up to one loop, whereas the other two terms are required at tree
level.

Taking all these issues in to account, we find the statistical entropy 
function up to order $Q^0$ as,
\ben
\Gamma^{stat}_{bmpv}(\vec \tau_B)&=& \Gamma_0(\vec \tau_B)+
\Gamma_1(\vec \tau_B)-\ln \hat G(\vec \tau_B)\nn
\Gamma_0 (\vec \tau_B) &=& {\Pi \over 2 \tau_{B_2}} |Q-\tau_BP|^2 \sim 
{\cal O}(Q^2)\nn
\Gamma_1 (\vec \tau_B) &=&  \ln g(\tau_B) +\ln g(-\bar\tau_B) 
+ (k+2) \ln (2\tau_{2B}) -\ln (4 \pi K_0) \sim {\cal O}(Q^0) \ . \nn
\een

\section{Correction to Statistical Entropy }
The statistical entropy of the system can be obtained by extremizing the
function $\Gamma^{stat}_{bmpv}$ and evaluating it at its extrema. It is an
straightforward exercise to check that for evaluating the entropy up to order
$Q^0$, it is enough to compute $\Gamma_{bmpv}$ at the extrema of $\Gamma_0$,
given as,
\be 
(\tau_0)_1={ Q\cdot P \over P^2} , \ \ \ \ \ \ (\tau_0)_2=
{\sqrt{Q^2P^2-(Q\cdot P)^2} 
\over P^2}.  
\ee 

Correction to this extrema due to $\Gamma_1$ will give ${\cal O}(Q^{-2})$
correction to the entropy. The expressions for the corrected entropy is,
\be
S^{stat}_{bmpv}=\Gamma^{stat}_{bmpv}(\vec \tau_0)\ .
\ee
Here, we give the approximate statistical entropies
$S^{(0)}_{stat}= S^{(0)} $ calculated using the `tree level' statistical
entropy function, $S^{(1)}_{stat}= S^{(0)}+S^{(1)} $ calculated using the
`tree level' and `one loop' statistical entropy function in a tabular form.

\begin{center}\def\st{\vrule height 3ex width 0ex}
\begin{tabular}{|l|l|l|l|l|l|l|} \hline 
$Q^2$ & $P^2$ & $Q\cdot P$ & $d(Q,P)$ & $S_{stat}$
& $S^{(0)}_{stat}$ & $S^{(1)}_{stat}$ 
\st\\[1ex]\hline \hline
2 & 2 & 0 &  $5424$ & 8.59 &  6.28
&  8.12  \st\\[1ex] \hline
4 & 4 & 0 &  $2540544$ & 14.74 &  12.57
&  14.40  \st\\[1ex] \hline
6 & 6 & 0 &  $1254480000$ & 20.95 &  18.85
&  20.69   \st\\[1ex] \hline
6 & 6 & 1 &  $991591800$ & 20.71 &  18.59
&  20.46   \st\\[1ex] \hline
6 & 6 & 2 &  $483665920$ & 20.00 &  17.77
&  19.76   \st\\[1ex] \hline
6 & 6 & -1 &  $991591800$ & 20.71 &  18.59
&  20.46   \st\\[1ex] \hline
6 & 6 & -2 &  $483665920$ & 20.00 &  17.77 
& 19.76  \st\\[1ex] \hline \hline
\end{tabular} 
\end{center}
We find that the asymptotic expansion of the statistical entropy is in good
agreement with the exact entropy of the system. The agreement is better for
higher values of charges. This is expected because asymptotic
expansion accurate for large charges but starts deviating for small
values of charges.

\section{Degeneracy for More General 5D Black Holes} 

The above analysis can easily be generalized to all five-dimensional CHL
black holes (for 4D CHL dyonic black holes see \cite{review}). These
are black
holes in the theory obtained by compactifying Heterotic and type IIB string
theory compactified on ${T^4 \times S^1 \over \mathbb{Z}_N}$, where
the $\mathbb{Z}_N$ group involves ${1 \over N}$ units of shift along the
$S^1$ circle and an order $N$ transformation on $T^4$.  This
transformation is chosen such that the theory preserves $\cal N=$$4$
supersymmetry. The partition function of these dyons is given by,
\be
f(\tilde \rho,\tilde \sigma,\tilde v )= e^{-2\pi i (\tilde \alpha \tilde\rho 
+ \tilde v)}
 \prod_{b=0}^1\, \prod_{r=0}^{N-1}
\prod_{k'\in \zzz+{r\over N},l\in \zzz, j\in 2\zzz+b
\atop k',l\ge 0, j<0 \, {\rm for}
\, k'=l=0}
\left( 1 - e^{2\pi i (\tilde \sigma k'   + \tilde \rho l + \tilde v
    j)}\right)^{
-\sum_{s=0}^{N-1} e^{-2\pi i sl/N } c_b^{(r,s)}(4lk' - j^2)}
\ee
where, $c_b^{(r,s)}(4lk' - j^2)$ are some constants that can be
obtained from the elliptic genus of the theory. Here also we can eliminate
the contribution from the KK-monopole sector and get the degeneracy
function for the generic five-dimensional black holes as,
\begin{eqnarray}
\tilde \Phi_{bmpv}(\tilde \rho, \tilde \sigma,\tilde v)&=&
{\tilde \Phi(\tilde \rho, \tilde \sigma,\tilde v) \over 
G(\tilde \rho, \tilde v)}, 
\een
where,
\ben \label{defg} 
  G(\tilde \rho, \tilde v)=64 e^{2 \pi i (\tilde \alpha \tilde \rho+
\tilde v)} (1-e^{-2 \pi i \tilde v})^2 \prod_{n=1}^{\infty}(1&-&e^{2 \pi i 
n \tilde \rho})^{-\sum_{s=0}^{N-1} 
e^{-2\pi i l s/N}
\, c^{(0,s)}_0(0)} \nn
(1-e^{2 \pi i(n \tilde \rho + \tilde v)})^{-\sum_{s=0}^{N-1} 
e^{-2\pi i l s/N}
\, c^{(0,s)}_1(-1)} 
(1&-&e^{2 \pi i(n \tilde \rho - \tilde v)})^{-\sum_{s=0}^{N-1} 
e^{-2\pi i l s/N}
\, c^{(0,s)}_1(-1)}.
\end{eqnarray}
With these modified expressions, one can proceed to compute first
subleading correction to the entropy of these general black
holes. For these orbifolded theories, the rank of the gauge group $r$
reduces and accordingly the number $k$ defined in (\ref{erank})
changes. Our previous analysis, corresponding to $r=28$ and $k=10$
goes through in all these cases. One can also produce an explicit chart
for systematic corrections to statistical entropy as we have in the 
previous section for these black holes, while there are quantitative
changes, qualitative behaviour remains the same.

\section{Discussion}
We studied the four-derivative $(O(Q^0))$ correction to the
statistical entropy function and the statistical entropy by doing
asymptotic expansion of the statistical entropy function.  This
expansion is valid in the limit (\ref{limit}), but is different from the
Cardy limit (or Fareytail limit \cite{faret1},\cite{faret2}), in our case
$Q^2 (=n)$ and $P^2 (Q_1Q_5)$ can be of same order whereas the Cardy
limit corresponds to $n >> Q_1Q_5$.

We find that the exact statistical degeneracy computed around the
saddle point $v=0$, is independent of the sign of $Q\cdot P$. It is
worthwhile to compare this with the four-dimensional black holes. In
4D case, the exact degeneracy changes as the sign of $Q\cdot P$
change. This jump in the degeneracy is related to the issue of walls
of marginal stability as discussed in details in \cite{walls1},
\cite{walls2}. As pointed out below (\ref{efo2}), there is an extra
zero at $\tilde v=0$ in $\tilde\Phi$, compared to
$\tilde\Phi_{bmpv}$. Because of this pole, there is an ambiguity in
the Fourier expansion and we get the jump in degeneracy for two signs
of $Q\cdot P$. Physically, it is related to the dynamics of the
KK-monopole. However, this sector is absent in the 5D BMPV black
holes.  For this five dimensional black holes, we do not have
any walls of marginal stability associated with this particular zero
of the function $\tilde\Phi_{bmpv}$.

\vspace{1cm}

\noindent
{\large {\bf Acknowledgement}}\\ 

We would like to thank Ashoke Sen for
suggesting this problem. We would also like to thank Dileep Jatkar and Ashoke
Sen for many useful discussions and their valuable comment on this draft. We
also thank Jyotirmoy Bhattecharya, Suvankar Dutta and Sameer Murthy for
discussions. We acknowledge the hospitality of Monsoon workshop on string
theory at TIFR, Mumbai, where the work was partially done.

\end{document}